# A Fast Convergence Density Evolution Algorithm for Optimal Rate LDPC Codes in BEC

H. Tavakoli, Assistant Professor
Department of Electrical
Engineering, Faculty of Engineering
University of Guilan, Rasht, Iran
htavakoli@guilan.ac.ir

*Abstract*—We derive a new fast convergent Density Evolution algorithm for finding optimal rate Low-Density Parity-Check (LDPC) codes used over the binary erasure channel (BEC). The fast convergence property comes from the modified Density Evolution (DE), a numerical method for analyzing the behavior of iterative decoding convergence of a LDPC code. We have used the method of [16] for designing of a LDPC code with optimal rate. This has been done for a given parity check node degree distribution, erasure probability and specified DE constraint. The fast behavior of DE and found optimal rate with this method compare with the previous DE constraint.

## I. Introduction

It is well known that for a reliable transmitting in a communication link, we need a proper coding and an efficient decoding algorithm. The Low-Density Parity-Check (LDPC) codes were first introduced by Gallager [1] and later were rediscovered by MacKay's [2]. This code is effective because of two important properties: achieving and approaching to the Shannon capacity of the channel in some circumstance [3]. Some works to achieve the channel capacity in infinite node degree distribution have been done in [4-6]. Moreover, infinite and finite length considerations for LDPC codes over Binary Erasure Channel (BEC) have been studied in [7-11].

An important characteristic for evaluating a LDPC code is successively decreasing error rate in decoding process by using a basic decoder such as message passing. Density Evolution (DE) is also used for analyzing the behavior of iterative message passing decoder in an infinite number of iterations. In fact, the behavior of DE for a code shows how the probability of successful decoding increases during decoding process. Due to this fact, every code has to satisfy the DE constraint in order to be applied in LDPC iterative decoders. This constraint is the most difficult part of code design problem.

One of the most popular communication models is BEC introduced by Elias in [12] and has the simplest DE constraint. This characteristic makes the BEC as a test channel for designing a code which it's rate can achieve and approach the channel capacity [13].

DE constraint in BEC was introduced and derived by Richardson and Urbanke in [3] and later in [14-15]. This constraint is a simple formula that illustrates the behavior of convergence in message passing decoding process. It means that, for any transmitted codeword in BEC, we can form the recursive iterations of the DE formula and follow the behavior of decreasing the erased probability in each iteration.

In this paper, we introduce a group of codes with optimal rate and fast convergence behavior in decreasing the erased probability in compared with those in [16-17]. We first review the convergence behavior of the code in BEC by DE formula. Then, some convergence properties in DE and related formulas are presented. An SDP reformulation of the fast convergence optimal rate problem is provided. Finally, the simulation results and concluding remarks complete the paper.

The rest of the paper is organized as follows: in Section 2, convergence of DE for considered definitions and the related formulas are illustrated. In Section 3, solution method of the optimal rate problem is considered. Section 4 provides a fast convergence approach for solving optimal rate problem. Finally, in Section 5 and Section 6, simulation results and conclusion finish the paper.

## II. Density Evolution: Definition, Formula and Convergence

Let H be a parity check matrix of a LDPC code with maximum variable node degree $D_v$ and check node degree $D_c$. A code can be defined by two polynomials as follows:

$$\rho(x) = \sum_{j=2}^{D_c} \rho_j x^{j-1} \qquad \lambda(x) = \sum_{i=2}^{D_v} \lambda_i x^{i-1} \qquad (1)$$

In both polynomials, coefficients, $\rho_j$ and $\lambda_i$, denote the probability of having a variable or check node with their related indices degree. Thus, we have

$$\sum_{j=2}^{D_c} \rho_j = 1 \qquad \sum_{i=2}^{D_v} \lambda_i = 1 \qquad \lambda_i \geq 0, \rho_j \geq 0 \qquad (2)$$

The related design rate of the code is defined as follows [15]:

$$R = 1 - \frac{\sum_{j=2}^{D_c} \rho_j/j}{\sum_{i=2}^{D_v} \lambda_i/i} \qquad (3)$$

For an erasure channel with erasure probability $\varepsilon > 0$ and given degree distributions, the necessary and sufficient condition for achieving the zero error probability in an infinite number of iterations in message passing decoder, i.e. DE constraint, is [3]:

$$\lambda(1 - \rho(1 - x)) \leq \frac{x}{\varepsilon} \quad \forall x \in [0, \varepsilon]. \qquad (4)$$

According to Massage Passing (MP) decoder of LDPC codes, for a proper code after each iteration, the bit error rate probability decreases. Using the definition of degree distribution of an LDPC code, this phenomenon is formulated as the so-called DE constraint. DE is a constraint for a good code that guarantees convergence of a code under MP decoding process.

Let us focus on the channel BEC and describe some related concepts. We start with some definitions.

**Definition 1 (DE constraint for BEC [7]):**

Let x and ε denote the probabilities of erasing a bit in l-th iteration and initial step of MP decoding process, respectively. Using Bays' rule, the probability of erasing a bit in $(l + 1)$-th iteration of MP decoding is defined by [3]:

$$y = \varepsilon \lambda(1 - \rho(1 - x)) \qquad (5)$$

In fact, y denotes the probability that a message node erased after one iteration of MP decoder from variable node to check node and from check node to variable node. The MP decoder converges if and only if :

$$y \leq x \quad \forall x \in [0, \varepsilon] \qquad (6)$$

**Definition 2 (Weak behavior of MP decoder):**

If degree distributions of a code satisfy the following constraints:

$$\varepsilon \lambda(1 - \rho(1 - x)) \leq x, \quad \sum \lambda_i = 1, \quad \sum \rho_j = 1,$$
$$0 \leq \varepsilon \leq 1, \quad 0 \leq x \leq \varepsilon \qquad (7)$$

then, $y \leq x$ is satisfied. In this case, we say that the MP decoder has weak behavior in this code.

**Definition 3 (Constraints of fast convergence DE):**

For a given $0 < \alpha < 1$, if degree distributions of a code satisfy the following constraints:

$$\varepsilon \lambda(1 - \rho(1 - x)) \leq \alpha x, \quad \sum \lambda_i = 1, \quad \sum \rho_j = 1,$$
$$0 \leq \varepsilon \leq 1, \quad 0 \leq x \leq \varepsilon \qquad (8)$$

then, $y \leq \alpha x$ is satisfied, according to (8). It is clear that α can not be zero. In this case, we say that the MP decoder has a fast convergence behavior in this code.

## III. OPTIMAL RATE PROBLEM AND SOLVING METHODS

In this section, we first state the optimal rate problem for approaching to the channel capacity as a semi-infinite optimization problem and then describe two methods for solving the problem, i.e., descretizing method and Semi-Definite reformulation.

### A. The Optimal Rate Problem

Let us consider the problem of finding the maximum achievable rate under the conditions that the check node degree and erasure probability are fixed. The main problem is to find good variable node degree to achieve the maximum rate

> NLP1: Max $\sum \frac{\lambda_i}{i}$
>
> Subject to: $\lambda_i \geq 0$
>
> $\sum \lambda_i = 1$
>
> $\lambda(1 - \rho(1 - \varepsilon x)) \leq x \quad \forall x \in [0,1]$

The latest constraint of the problem NLP1 is equivalent to (4). The problem NLP1 consists of infinite number of constraints and is a semi-infinite optimization problem. In the next section, we will describe the ways for solving NLP1.

### B. Algorithms for Solving Optimal Rate Problem

In [18] the authors classified the main approaches for finding good degree distributions. According to [18], there are two important ways for solving the problem NLP1. The first method is based on LP methods in which instead of considering the whole interval $(0,1]$, by choosing a set of sample points $\{x_0, x_1, \ldots, x_N\} \subseteq (0,1]$ for x, one discretizes the DE inequality to linear inequalities [8, Section V]. The LP problem related to the sample points $\{x_0, x_1, \ldots, x_N\}$ can be presented as follows:

> LP1: Max $\sum \frac{\lambda_i}{i}$
>
> Subject to: $\lambda_i \geq 0$
>
> $\sum \lambda_i = 1$
>
> $\lambda(1 - \rho(1 - \varepsilon x_i)) \leq x_i \quad 1 \leq i \leq N$

## IV. A FAST CONVERGENT APPROACH FOR SOLVING OPTIMAL RATE PROBLEM

In this section, in order to design a good code with the fast convergence property, we provide a new approach for speeding up the convergence process of the second approach mentioned in the Section 3. For this purpose, using the Definition 3 of DE constraint, we arrive to the following optimization problem which has fast convergence property:

> NLP2: Max $\sum \frac{\lambda_i}{i}$
>
> Subject to: $\lambda_i \geq 0$

$$\sum \lambda_i = 1$$
$$\lambda(1 - \rho(1 - \varepsilon x)) \leq \alpha x \quad \forall x \in [0,1]$$

For the given $\varepsilon$ and $\alpha$ in the interval $(0,1)$ and given parity check node degree distribution $\rho(x)$, using the similar approach in [18], the SDP problem related to the problem NLP2 can be described as follows:

SDP2: Max $\sum \frac{\lambda_i}{i}$

Subject to: $\sum \lambda_i = 1$

$\Pi_l = \sum_{i+j=l} B_{ij}, \quad 0 \leq l \leq n$

$B \succcurlyeq 0, \quad 0 \leq \lambda_i \leq 1.$

In this way, each elements of the vector $\Pi$ is a linear combination of $\lambda_i$s and $\alpha$. We also have to note that using different values for $\alpha$ leads us to different solutions of SDP2. In the next section, some simulation results are presented to show the efficiency and convergence rate of this approach in comparison with the approaches provided in [18].

## V. SIMULATION RESULTS

In this part of paper, we simulate the optimal rate problem for finding an optimal rate code for different values of $\alpha$. As it appears, to gain the fast convergence property, we lose the maximum optimal rate property. For providing Simulation results, for each value of $\alpha$, we solve the SDP2 and find variable degree distribution, which gives optimal rate. It is clear if $\alpha = 1$, the results of [16-17] found.

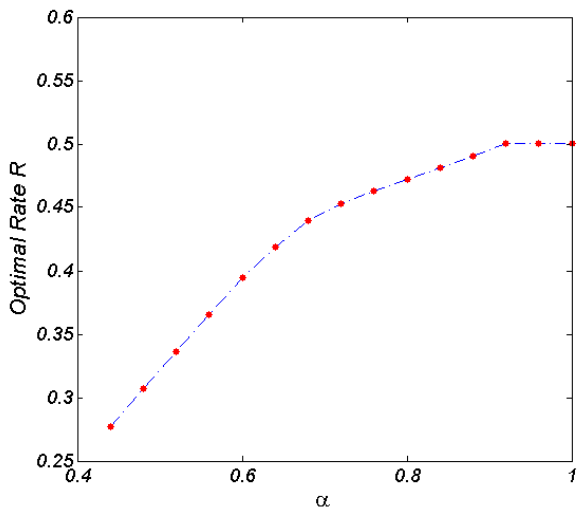

Fig.1: illustration of decreasing the optimal rate found by SDP method, vs. increasing $\alpha$ for $\rho(x) = x^3$, $\lambda(x) = \lambda_2 x + \lambda_3 x^2 + \lambda_4 x^3 + \lambda_5 x^4 + \lambda_6 x^5$ and $\varepsilon = 0.3$

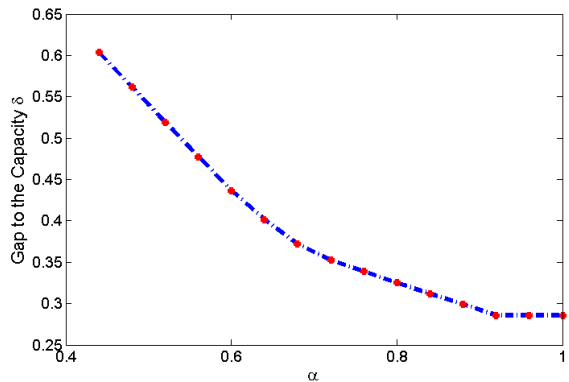

Fig.2: illustration of decreasing the gap to the capacity, defined as: $\delta = 1 - R/C$ vs. increasing $\alpha$ for $\rho(x) = x^3$, $\lambda(x) = \lambda_2 x + \lambda_3 x^2 + \lambda_4 x^3 + \lambda_5 x^4 + \lambda_6 x^5$ and $\varepsilon = 0.3$

## VI. CONCLUSION

In this paper, we reformulate DE inequality for deriving fast convergence codes. The reformulated DE inequality solved by SDP method and simulation result presented. Simulation results show that we have a trade-off between fast convergence codes and optimal rate. We can see the behavior of fast convergence in Fig.3 . The derived codes with fast convergence property can be effectively used in a network with several relays with full or partial Decode and Forward strategy.

## REFERENCES

[1] R. G. Gallager, Low-Density Parity-Check Codes. Cambridge, MA: MIT Press, 1963.

[2] D. J. C. MacKay and R. M. Neal, Near Shannon limit performance of low density parity-check codes, IEE Electronic Letters, vol. 32, no. 18, pp. 1545–1646, August 1996.

[3] T. Richardson, A. Shokrollahi and R. Urbanke, Design of capacity approaching irregular low-density parity-check codes, IEEE Trans. on IT, vol. 47, no. 2, pp. 619–637, February 2001.

[4] A. Shokrollahi, Capacity-achieving sequences, IMA Volume in Mathematics and its Applications, vol. 123, pp. 153–166, 2000.

[5] P. Oswald and A. Shokrollahi, Capacity-achieving sequences for the erasure channel, IEEE Trans. on IT, vol. 48, no. 12, pp. 3017–3028, December 2002.

[6] A. Shokrollahi, New sequences of linear time erasure codes approaching the channel capacity, in Proc. 13th ISAA, no. 1719 in LNCS, pp. 65-76, 1999.

[7] M. G. Luby, M. Mitzenmacher, M. A. Shokrollahi and D. A. Spielman, Efficient erasure-correcting codes, IEEE Trans. on IT, vol. 47, no. 2, pp. 569–584, February 2001.

[8] C. Di, D. Proietti, I. E. Telatar, T. J. Richardson, and R. L. Urbanke, Finite length analysis of low-density parity-check codes on the binary erasure channel, IEEE Trans. IT, vol. 48, no. 6, pp. 1570-1579, June 2002.

[9] T. J. Richrdson, A. Shokrollahi, and R. Urbanke, Finite-length analysis of various low-density parity-check ensembles for the binary erasure channel, in Proc. IEEE ISIT, Switzerland, July 2002,p. 1.

[10] D. J. Costello and G. D. Forney, Channel coding: The road to channel capacity, Proceedings of the IEEE, vol. 95, no. 6, pp. 1150–1177, June 2007.

[11] Special issue on Codes on Graphs and Iterative Algorithms, IEEE Trans. on IT, vol. 47, no. 2, February 2001.

[12] P. Elias. Coding for two noisy channels, 3rd London Symp. on IT, pp. 61–76, 1955.

[13] H. D. Pfister, I. Sason and R. Urbanke, Capacity-achieving ensembles for the binary erasure channel with bounded complexity, IEEE Trans. on IT, vol. 51, no. 7, pp. 2352–2379, July 2005.

[14] T. Richardson and R. Urbanke, The capacity of low-density parity-check codes under message-passing decoding, IEEE Trans. on IT, vol. 47, no. 2, pp. 599–618, February 2001.

[15] T. Richardson and R. Urbanke, Modern Coding Theory, Cambridge University Press, 2008. [Online]. Available: http://lthcwww.epfl.ch/mct/index.php.

[16] H.Tavakoli, M.Ahmadian Attari, M.R.Peyghami, Optimal Rate for Irregular LDPC Codes in Binary Erasure Channel, ITW 2011, Paraty, Brasil, October 16-20, 2011.

[17] H.Tavakoli, M.Ahmadian Attari, M.R.Peyghami, Optimal Rate and Maximum Erasure Probability LDPC Codes in Binary Erasure Channel, 49th Allerton Conference, September 28-30, 2011, Illinois, USA

[18] H.Tavakoli, M.Ahmadian Attari, M.R.Peyghami, Optimal Rate Irregular LDPC Codes in Binary Erasure Channel, submitted to IET communications